\documentclass[a4paper]{article}

\usepackage{graphicx}
\usepackage[latin10]{inputenc}
\usepackage[OT1]{fontenc}

\usepackage{algorithm}
\usepackage[noend]{algpseudocode}
\usepackage{amsfonts}
\usepackage{graphicx}
\usepackage{mymacros}
\usepackage{array}
\usepackage[defblank]{paralist} 

\usepackage{RR}
    \RRprojet{Regal}
    \RRtheme{\THCom}
    \URRocq

\begin{document}
\RRNo{6956}
\RRauthor{
    Mihai Le{\textcommabelow{t}}ia%
    \thanks{École Normale Supérieure de Lyon and LIP6}
    \and{}
    Nuno Pregui{\c c}a%
    \thanks{Universidade Nova de Lisboa}
    \and{}
    Marc Shapiro%
    \thanks{INRIA Paris-Rocquencourt and LIP6}
}
\authorhead{Le{\textcommabelow{t}}ia, Preguiça, Shapiro}

\RRdate{Juin 2009}

\RRtitle{Les CRDT\,: Cohérence sans contrôle de concurrence}
\titlehead{CRDTs: Consistency without concurrency control}
\RRetitle{CRDTs: Consistency without concurrency control%
\thanks{
    This work is supported in part by the EU FP6 project Grid4All, a PhD
    grant from Microsoft Research, and the
    Portuguese FCT/MCTES project POSC/59064/2004, with FEDER funding.
}
}

\RRabstract{
      A CRDT is a data type whose operations commute when
  they are concurrent.
  Replicas of a CRDT eventually converge without any complex concurrency
  control.
  As an existence proof, we exhibit a non-trivial CRDT: a shared edit
  buffer called Treedoc.
  We outline the design, implementation and performance of Treedoc.
  We discuss how the CRDT concept can be generalised, and its
  limitations.
}

\RRresume{
  Un CRDT est un type de données dont toutes les opérations concurrentes
  sont commutatives.
  Les répliques d'un CRDT convergent inéluctablement, sans nécessiter un
  contrôle de concurrence complexe.
  Comme preuve d'existence, nous montrons un CRDT non trivial\,: un
  tampon d'édition partagée appelé Treedoc.
  Nous en résumons la conception, la mise en {\oe}uvre et les
  performances.
  Nous discutons les limites et les possibles généralisations du concept.
}

\RRmotcle{Réplication des données, réplication optimiste, opérations
commutatives}
\RRkeyword{Data replication, optimistic replication, commutative
operations}

\makeRR

\section{Introduction}
\label{sec:introduction}

Shared read-only data is easy to scale by using well-understood
replication techniques.
However, sharing \emph{mutable} data at a large scale is a difficult
problem, because of the CAP impossibility result \cite{rep:pan:1628}.
Two approaches dominate in practice.
One ensures scalability by giving up  consistency guarantees, for
instance using the Last-Writer-Wins (LWW) approach
\cite{db:rep:optim:1454}.
The alternative guarantees consistency by serialising all updates,
which does not scale beyond a small cluster \cite{pro:pan:1627}.
Optimistic replication allows replicas to diverge, eventually
resolving conflicts either by LWW-like methods or by  serialisation
\cite{optim:rep:syn:1500}.

In some (limited) cases, a radical simplification is possible.
If concurrent updates to some datum commute, and all of its replicas
execute all updates in causal order, then the replicas converge.%
\footnote{
  Technically, LWW operations commute; however they achieve this by
  throwing away non-winning operations.
  We aim instead for \emph{genuine} commutativity that does not lose
  work, i.e., the output should reflect the cumulative effect of the
  operations.
  }
We call this a Commutative Replicated Data Type (CRDT).
The CRDT approach ensures that there are no conflicts, hence, no need
for consensus-based concurrency control.
CRDTs are not a universal solution, but, perhaps surprisingly, we
were able to design highly useful CRDTs.
This new research direction is promising as it ensures consistency in
the large scale at a low cost, at least for some applications.

A trivial example of a CRDT is a set with a single \emph{add-element}
operation.
A \emph{delete-element} operation can be emulated by adding
``deleted'' elements to a second set.
This suffices to implement a mailbox \cite{baquero}.
However, this is not practical, as the data structures grow without
bound.
A more interesting example is WOOT, a CRDT for concurrent editing
\cite{app:rep:1587}, pioneering but inefficient, and its successor
Logoot \cite{app:rep:1625}.

As an existence proof of non-trivial, useful, practical and efficient
CRDT, we exhibit one that implements an ordered set with
insert-at-position and delete operations.
It is called Treedoc, because sequence elements are identified compactly
using a naming tree, and because its first use was concurrent
document editing \cite{Preguica2009}.
Its design presents original solutions to scalability issues, namely
restructuring the tree without violating commutativity, supporting very
large and variable numbers of writable replicas, and leveraging the data
structure to ensure causal ordering without vector clocks.

Another non-trivial CRDT that we developed (but we do not describe here)
is a high-performance shared, distributed graph structure, the
\emph{multilog} \cite{rep:optim:sh126}.

While the advantages of commutativity are well documented, we are the
first (to our knowledge) to address the design of CRDTs.
In future work, we plan to explore what other interesting CRDTs may
exist, and what are the theoretical and practical requirements for
CRDTs.

The contributions of this paper are the following:
\begin{inparablank}
\item
  We exhibit a non-trivial, practical, efficient CRDT\@.
\item
  We address practical issues in CRDT design such as indefinite growth,
  identifier size, restructuring and garbage collection.
\item
  We present a novel approach side-stepping the non-scalability of
  consensus when dealing with dynamic, varying numbers
  of sites.
\item
  We present some experimental data based on Wikipedia traces.
\end{inparablank}

The paper proceeds as follows.
This introduction is Section~\ref{sec:introduction}.
Section~\ref{sec:ordered-sequence} presents our ordered-sequence CRDT
abstraction.
Section~\ref{sec:performance} examines the trace data and experimental
performance of our CRDT.
In Section~\ref{sec:treedoc-LS} we present our solutions to some
specific scalability issues.
Section~\ref{sec:strengths+weaknesses} discusses lessons learned and
possible generalisations.
Section~\ref{sec:conclusion} concludes and outlines future work.

\section{An ordered-set CRDT}
\label{sec:ordered-sequence}

We begin by considering the requirements of a CRDT providing the
abstraction of an ordered sequence of (opaque) \emph{atoms}.

\subsection{Model}
\label{sec:model}

We consider a collection of \emph{sites} (i.e., networked computers),
each carrying a \emph{replica} of a shared ordered-set object, and
connected by a reliable broadcast protocol (e.g., epidemic
communication).
We support a peer-to-peer, multi-master execution model: some arbitrary
site \emph{initiates} an update and executes it against its local
replica; each other site eventually receives the operation and
\emph{replays} it against its own replica.
All sites eventually receive and execute all operations;
causally-related operations execute in order, but concurrent operations
may execute in different orders at different sites.

The update operations of the ordered-set abstraction are the following:
\begin{itemize}
\item
  \emph{insert($\ID$,newatom)}, where $\ID$ is a fresh identifier.
  This operation adds atom \emph{newatom} to the ordered-set.
\item
  \emph{delete($\ID$)}, deletes the atom identified $\ID$ from the
  ordered-set.
\end{itemize}

Two inserts or deletes that refer to different IDs commute.
Furthermore, operations are idempotent, i.e., inserting or deleting the
same ID any number of times has the same effect as once.
To ensure commutativity of concurrent inserts, we only need to ensure
that no two IDs are equal across sites.
Our ID allocation mechanism will be described next.

\subsection{Identifiers}
\label{sec:tid-requirements}

Atom identifiers must have the following properties:
\begin{inparaenum}[\it (i)]
\item
  Two replicas of the same atom (in different replicas of the
  ordered-set) have the same  identifier.
\item
  No two atoms have the same identifier.
\item \label{item:constant}
  An atom's identifier remains constant for the entire lifetime of the
  ordered-set.%
\footnote{
  Later in this paper we will weaken this property.
}
\item
  There is a total order ``$<$'' over identifiers, which
  defines the ordering of the atoms in the ordered-set.
\item \label{item:dense}
  The identifier space is \emph{dense}.
\end{inparaenum}

Property~(\ref{item:dense}) means that between any two identifiers $P$ and
$F$, $P<F$, we can allocate a fresh identifier $N$, such that
$P<N<F$.
Thus, we are able to insert a new atom between any two existing ones.

The set of real numbers $\mathbb{R}$ is dense, but cannot be used for
our purpose, because, as atoms are inserted, the precision required will
grow without bound.
We outline a simpler solution next.

\begin{figure}[t]
 \begin{center}
    \includegraphics[height=2.2cm,bb=92 600 346 746]{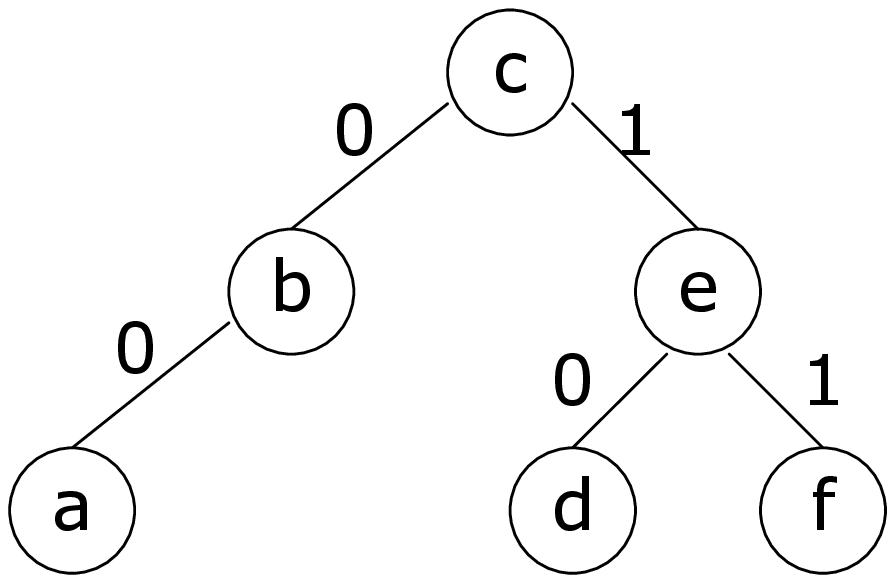}
  \end{center}
  \caption{Example Treedoc.  The TID for \emph{"b"} is 0; the TID of
    \emph{"c"} is the empty string; the TID of \emph{"d"} is 10.}
  \label{fig:simple}
\end{figure}
\begin{figure}[t]
  \begin{center}
    \includegraphics[height=1.2cm,bb=92 700 320 764]{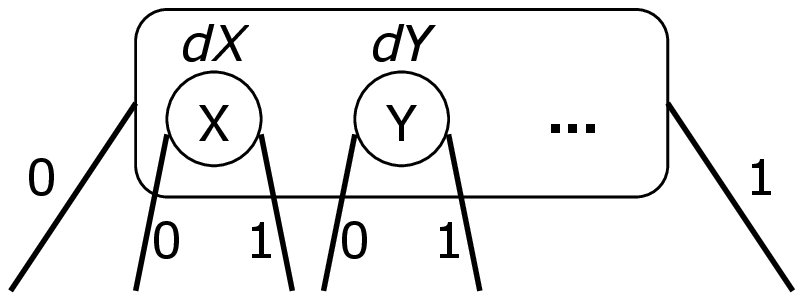}
  \end{center}
  \caption{A treedoc major node}
  \label{fig:major}
\end{figure}
\begin{figure}[t]
  \begin{center}
    \includegraphics[height=2.5cm,bb=92 600 338 762]{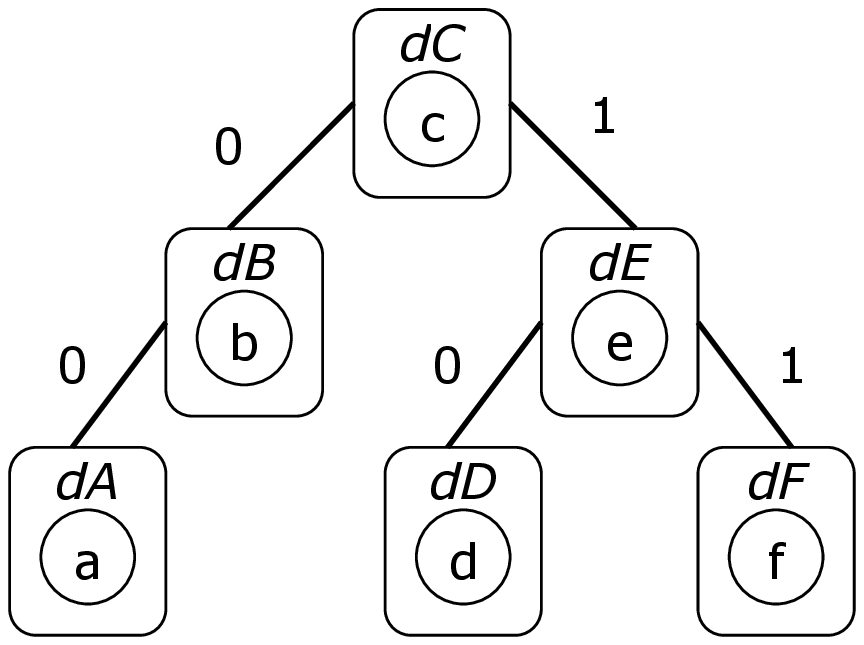}
  \end{center}
  \caption{A treedoc node with disambiguators}
  \label{fig:disamb}
\end{figure}

\subsection{The Treedoc CRDT}
\label{sec:tid}
\label{sec:disambiguator}

In Treedoc, an atom identifier, henceforth called a TID, represents a
path in a tree.
If the tree is balanced, the average TID size is logarithmic in the
number of atoms.
We experimented with both binary and 256-ary trees; for lack of space we
present only the binary version.
The order ``$<$'' is infix traversal order (i.e., left to right).
Figure~\ref{fig:simple} shows a binary Treedoc that contains the text
\emph{"abcdef"}.

In a distributed environment, different sites might concurrently
allocate the same TID.
To avoid this, we extend the basic tree structure, allowing a node to
contain a number of internal nodes, called \emph{mini-nodes}.
A node containing mini-nodes will be called a \emph{major node}.
Figure~\ref{fig:major} shows an example major node.
Inside a major node, mini-nodes are distinguished by a
\emph{disambiguator} that identifies the site that inserted the node.
Disambiguators are unique and ordered, giving a total order between
entries in the ordered-set.

Figure~\ref{fig:disamb} shows a Treedoc structure with disambiguators
represented at every node.
Site A with disambiguator $dA$ inserted atom $a$, site B inserted atom
$b$, and so on.
Mini-nodes are traversed in disambiguator order.




\subsection{Treedoc insert and delete}

We now describe the ordered-set update operations, insert and delete.
We start with delete, the simpler of the two.
A $delete(\TID)$ simply discards the atom associated with $\TID$.
We retain the corresponding tree node and mark it as a \emph{tombstone}.
(In certain cases, out of the scope of this short paper, a
tombstone may be discarded immediately.)

To insert an atom, the initiator site chooses a fresh TID that positions
it as desired relative to the other atoms.
For instance, to insert an atom $R$ to the right of atom $L$:
\begin{inparaitem}
\item
  If $L$ does not have a right child, the TID of $R$ is the TID of $L$
  concatenated with 1 ($R$ becomes the right child of $L$).
\item
  Otherwise, if $L$ has a right child $Q$, then allocate the TID of the
  leftmost position of the subtree rooted at $Q$.
\end{inparaitem}

\subsection{Restructuring the tree}
\label{sec:flatten}

In the approach described so far, depending on the pattern of inserts
and deletes, the tree may become badly unbalanced or riddled with
tombstones.
To alleviate this problem, the new restructuring operation
\emph{flatten} transforms a tree into a flat array,
eliminating all storage overhead.
As a flat array can equivalently be interpreted as a balanced tree,
there is no need for the inverse operation.
As the flattening operation changes the TIDs, we modify
Property~(\ref{item:constant}) of Section~\ref{sec:tid-requirements} to
allow non-ambiguous renaming.

However, flattening does not genuinely commute with update operations.
We solve this using an update-wins approach: if a flatten occurs
concurrently with an update, the update wins, and the flatten aborts
with no effect.
We use a two-phase commit protocol for this purpose (or, better, a
fault-tolerant variant such as Paxos Commit \cite{db:syn:1578}).
The site that initiates the flatten acts as the coordinator and collects
the votes of all other sites.
Any site that detects an update concurrent to the flatten votes ``no'',
otherwise it votes ``yes.''
The coordinator aborts the flatten if any site voted ``no'' or if some
site is crashed.
Commitment protocols are problematic in large-scale and dynamic systems;
in Section~\ref{sec:treedoc-LS} we explain how we solve this issue.




\begin{figure}[t]
 \begin{center}
    \includegraphics{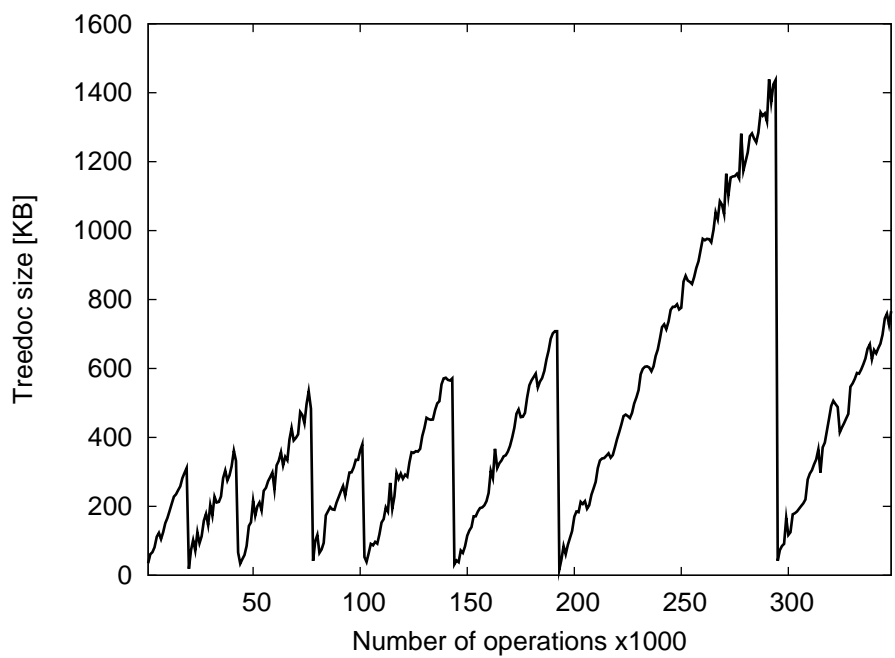} 
  \end{center}
  \caption{Treedoc size over time (GWB page)}
  \label{fig:size}
\end{figure}
\begin{figure}[t]
 \begin{center}
    \includegraphics{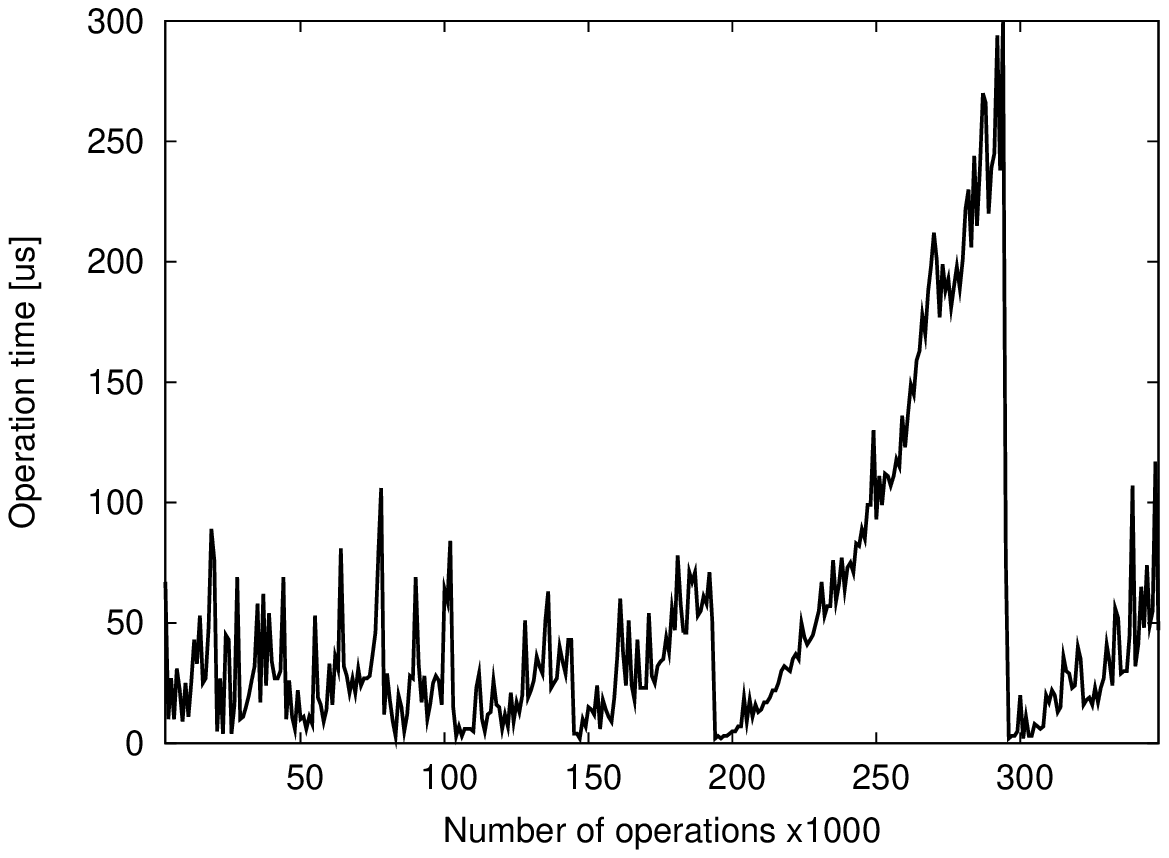} 
  \end{center}
  \caption{Execution time per operation (GWB page)}
  \label{fig:time_flat}
\end{figure}

\section{Experiments}
\label{sec:performance}

We ran a series of experiments based on cooperative editing traces.

A number of Wikipedia pages were stored as Treedocs, interpreting
differences between successive versions of a page as a series of inserts
and deletes.
In some experiments our atoms were words; in the ones reported here an
atom is a whole paragraph.
We also ran similar experiments based on traces of SVN repositories
containing LaTeX{} Java source code.
A common observation across all experiments is that the number of
deletes is surprisingly high.

We studied medium-sized Wikipedia pages such as ``Distributed
Computing,'' reaching 20\,KB of text in 800 revisions, or ``PowerPC''
reaching 25\,KB in 400 revisions.
Applying all the revisions for these pages required less than 1 second
when using paragraphs as atoms, and 2 seconds using words.
We also studied some frequently-edited pages, e.g., ``George W.
Bush'' (GWB) reaching 150\,KB in $40,000$ revisions.
Because of vandalism, the GWB page contains an even higher proportion of
deletes (in the absence of flattening, 95\% of nodes would be
tombstones).

Hereafter we report only on the most stressful benchmark, i.e., the GWB
traces, with a 256-ary tree, and full paragraphs as atoms, flattening
every $1,000$ revisions; $1,000$ successive revisions may include up to
$100,000$ update operations.


Figure~\ref{fig:size} shows the size of the GWB Treedoc structure over
the first $350,000$ edit operations of the GWB page.
Size increases with successive edits, then falls dramatically at each
periodic flattening.
The decrease is attributable mostly to discarding tombstones, but also
to improved balance of the tree: thus, the average TID size shrinks from
60 bytes before flattening to only 2 bytes.

Figure~\ref{fig:time_flat} shows execution time per operation.
Again, flattening has dramatic effect.
Without flattening, the per-operation time would grow up to 3\,ms.
Periodic flattening decreases the depth of the tree to
2-3 levels, and the slowest update takes only $0.3\,ms$.

From this we can estimate the scalability of Treedoc for
concurrent updating.
Assume that every site continuously initiates one update every 3 seconds.
Then the system can sustain $1,000$ simultaneous users without
flattening, and $10,000$ when flattening at $1,000$-revision intervals.

\section{Treedoc in the large scale}
\label{sec:treedoc-LS}

The CRDT approach guarantees that replicas converge.
However, we saw that metadata accumulates over time, with a big impact
on performance, and must be garbage-collected from time to time.
The attendant commitment or consensus is a problem for scalability.
In this section, we explain how Treedoc addresses this issue.

\subsection{Supporting large and dynamic numbers of replicas}
\label{sec:core+nebula}

Commitment requires the unanimous agreement of a
well-identified set of sites.
This is problematic in a large-scale system, where sites
fluctuate dynamically.
In scenarios such as collaborative editing, new participants may enter
at any time, leave the system definitely, or disconnect for an
unpredictable period of time, while continuing to initiate updates.

To solve this problem, we partition the sites in two disjoint subsets.
The \emph{core} consists of a small group of sites that are well-known
and well-connected.
In the limit, the core could reduce to a single server.
The sites that are not in the core are part of the \emph{nebula}.
Only core sites participate in commitment.

\subsection{Nebula catch-up protocol}
\label{sec:catch-up}

Let us call an interval between successful flattens an \emph{epoch}.
Each flatten -- each change of epoch -- changes the frame of reference
for TIDs: TIDs from some epoch are invalid in other epochs, and sites
may exchange updates only if in the same epoch.
Core sites are in the same epoch by construction, but a nebula site may
lag behind.

In order to communicate with the nebula, a core site executes a
\emph{catch-up} protocol, which we now describe at a high level.
To simplify the description, assume that the core and the nebula sites
started from the same initial state, and that the core executed a single
flatten since then: If the core is in epoch $n$ (the ``new'' epoch), the
nebula is in epoch $n-1$ (``old'' epoch).
Updates in the old epoch use ``old'' TIDs, whereas those in the new
epoch use ``new'' TIDs.

A core site maintains a buffer of update messages that it needs to send
to the nebula, some in the old epoch, some in the new one.
Conversely, a nebula site maintains a buffer of update messages to be
sent to the core; they are all in the old epoch.

Old messages buffered in the core can be sent to the nebula site
(operating in the old epoch) and replayed there.
However, the converse is not true: since the core is in the new epoch,
it cannot replay old updates from the nebula.
The nebula must first bring them into the new epoch.
To this effect, and once it has applied all old core updates, the nebula
site flattens its local replica of the tree, using the tree itself to
keep track of the mapping between old and new TIDs.
Then it translates old TIDs in buffered messages into the corresponding
new TIDs.
At this point, the nebula site is in the new epoch.
(It may now either join the core, or remain in the nebula.)
Finally, it sends its buffered messages to the core, which can now
replay them.

Since epochs are totally ordered, and since every nebula site will go
through the same catch-up protocol, concurrent updates remain
commutative, even if initiated in different epochs.

\subsection{Core/nebula requirements}
\label{sec:core-neb-requirements}

The requirements for managing the core and nebula follow from the above
description.

Joining or leaving the core follows a membership protocol
\cite{rep:pan:1629}.
All core sites participate in flattening.
Core sites may freely initiate updates and replay each others'.

Sites in the nebula are assumed to be uniquely identified (for
disambiguators), but are otherwise unrestricted.
The nebula may contain any number of sites, which are connected to the
network or disconnected.
Nebula sites may freely initiate updates, but do not participate in
commitment.

Two sites may send updates to each other, and replay each others' updates,
if and only if they are in the same epoch.

\subsection{TID translation algorithm}
\label{sec:catch-up-protocol}

We now describe in more detail how a nebula site translates TIDs from
the old to the new epoch.
It needs to distinguish operations that were received from the core and
are serialised before the flatten, from those initiated locally or
received from other nebula sites, which must be serialised after the
flatten.
For this purpose we colour the corresponding nodes either Cyan (C for
Core) or Black (Noir in French, N for Nebula).

Thus we distinguish cyan nodes, cyan tombstones, black nodes and black
tombstones.
A node can be both a cyan node and a black tombstone; the converse is
not possible.

We will now describe the steps that a nebula site needs to take in order to execute
a flatten operation. We will assume that all the operations from the core issued prior
to the flatten have been executed as well as some black operations,
some local and some from other nebula sites.
Once the flatten is performed the site will be able to send the black operations to the core.
The flatten will construct list of subtrees, each having as root a cyan node.

The first step is to go through the tree and examine only cyan nodes and tombstones.
We ensure that a sentinel node $n_b$ always exists to mark the beginning of the ordered-set and to
ensure the tree is not empty. We identify the following cases:
\begin{itemize}
  \item \textbf{cyan node} (can also be a black tombstone) - add to the list
    along with any black children it has
  \item \textbf{cyan tombstone} - add any black children to the subtree of the last
    node in the flattened list.
    We preserve the correct order by adding at the end of the subtree.
    If no cyan nodes have been seen so far, we add the black children
    to $n_b$.
\end{itemize}

The second step is to create the new balanced tree from the roots of the subtrees
stored in the linear list. The nodes that have black children will be transformed
into major nodes if both a cyan child and a black child should be placed on the
same position.

The last step is to go though the new tree and generate the update operations to be sent
to the core. We examine only black nodes and tombstones:
\begin{itemize}
  \item \textbf{black node} - send \emph{insert} operation with this TID{} and atom
  \item \textbf{black tombstone} - send \emph{delete} operation with this TID
\end{itemize}

When a nebula site connects to the core,
it sends not only black operations generated locally,
but also operations received from other nebula sites.
It may happen that a site receives the same update multiple times,
but this causes no harm since updates are idempotent.

\subsection{Approximate causal ordering}

Vector clocks are commonly used to ensure causal ordering and to suppress
duplicate messages.
We observe that causal ordering is already encoded in the Treedoc
structure: inserting some node always happens-before inserting some
descendant of that node, and always happens-before the deletion of the
same node.
Operations on unrelated nodes, even if one happened-before the other,
can execute in any order.
Furthermore, duplicate messages are inefficient but cause no harm, since
operations are idempotent.
Therefore, a precise vector clock is not necessary; approximate variants
that are scalable may be sufficient as long as they
suppress a good proportion of duplicates.










\section{Discussion}
\label{sec:strengths+weaknesses}

Massive distributed computing environments, such as Zookeeper or Dynamo
\cite{app:rep:optim:1606}, replicate data to achieve high availability,
performance and durability.
Achieving strong consistency in such environments is inherently
difficult and requires a non-scalable consensus; however in the absence
of consistency, application programmers are faced with overwhelming
complexity.
For some applications, eventual consistency is sufficient
\cite{app:rep:optim:1606}, but complexity and consensus are hiding under a
different guise, namely of conflict detection and resolution.

In this paper we propose to use CRDTs because they ensure eventual
consistency without requiring consensus.
Although garbage collection is based on consensus, it remains outside
the critical path of the application and hidden inside the abstraction
boundary.

Not all abstractions can be converted into a CRDT: for instance a queue
or a stack rely on a strong invariant (a total order) that inherently
requires consensus.
Treedoc on the other hand maintains a local, partial order, and the
outcome of its operations need not be unique.

Even when an abstraction is not a CRDT, it is very useful to design it
so that most pairs of operations commute when concurrent.
Those pairs can benefit from cheap, high-performance protocols,
resorting to consensus only for non-commuting pairs \cite{rep:syn:1567}.

Generalising from Treedoc and Multilog (see Introduction) teaches us a
few interesting lessons about the requirements for CRDTs.
To commute, operations must have identical pre-condition; in practice,
all operations should have pre-condition ``true.''
A central requirement is the use of unique, unchanging identifiers.
To be practical, the data structure must remain compact; we ensure this
by using an ever-growing tree, ensuring that metadata and identifiers remain
compact (logarithmic in the size of the data).

\section{Conclusion}
\label{sec:conclusion}

The Commutative Replicated Data Type or CRDT is designed to make
concurrent operations commute.
This removes the need for complex concurrency control allowing
operations to be executed in different orders and still have replicas
converge to the same result.
CRDTs enable increased performance and scalability compared to classical
approaches.

Although designing a CRDT to satisfy certain requirements is not always
possible, loosening invariants or precision constraints should allow the
design of commutative operations.

We have proposed a CRDT called Treedoc that maintains an ordered set of
atoms while providing insert and delete operations.
To overcome the challenges of practicality and scalability, we explored
some innovative solutions.
Each atom has a unique, system-wide, compact identification that does
not change between flattens.
Garbage collection is a requirement in practice; it is disruptive and
requires consensus, but it has lower precedence that updates, and it is
not in the critical path of applications.
We side-step the non-scalability of consensus by dividing sites into two
categories with different roles.
CRDTs require causal ordering, but since the Treedoc metadata encodes causal
ordering implicitly, it does not need to be maintained strictly at
the system level; this enables the use of scalable approximations of
vector clocks.

Our future work includes searching for other CRDTs as well as studying
the interaction between CRDTs and classical data structures.

\bibliographystyle{acm}
\bibliography{bib,shapiro-bib,licenta}
\end{document}